%% file: infonet.tex
\newif\ifnips
\nipstrue

\ifnips
\documentclass[10pt]{article}
\usepackage{nips2004e,times,amsmath,amssymb,amsfonts}
\else
\documentclass[aps,twocolumn,
amsmath,amsfonts,final,floatfix,nofootinbib
]{revtex4}
\fi

\newif\ifpdf 
\ifx\pdfoutput\undefined
\pdffalse 
\else
\pdfoutput=1 
\pdftrue
\fi

\ifpdf 
\usepackage[pdftex]{graphicx}
\else 
\usepackage{graphicx}
\fi

\newcommand{\nipsprl}[2]{\ifnips #1\else #2\fi}
\usepackage[cdot,squaren,textstyle]{SIunits}
\nipsprl{
  \usepackage[sort&compress,numbers]{natbib}
  \nipsprl{\setlength{\bibsep}{0mm}}{}
}{
  \usepackage[sort&compress]{natbib}
}
\usepackage{url}
\usepackage{floatflt}

\input{aliases}

\newtheorem{theorem}{Theorem}

\begin{document}

\title{Information theory, multivariate dependence, and genetic
  network inference}

\nipsprl{
\author{Ilya Nemenman\\Kavli Institute for Theoretical Physics, University of
  California, Santa Barbara, CA 93106\\\url{nemenman@kitp.ucsb.edu}}
}{
\author{Ilya Nemenman}
\email{nemenman@kitp.ucsb.edu}
\affiliation{Kavli Institute for Theoretical Physics, University of
  California, Santa Barbara, California 93106}
}

\nipsprl{\maketitle}{}
\begin{abstract}
  We define the concept of dependence among multiple variables using
  maximum entropy techniques and introduce a graphical notation to
  denote the dependencies. Direct inference of information theoretic
  quantities from data uncovers dependencies even in undersampled
  regimes when the joint probability distribution cannot be reliably
  estimated. The method is tested on synthetic data. We anticipate it
  to be useful for inference of genetic circuits and other biological
  signaling networks.
\end{abstract}
\nipsprl{}{\pacs{}
  \preprint{}
  \maketitle
}

\section{Two problems}
One of the most active fields in quantitative biology is the inference
of biological interaction networks (e.\ g., protein or genetic
regulatory networks) from high throughput data such as expression
microarrays \cite{friedman-04}\footnote{The literature on reverse
  engineering biological networks develops quickly, and we do not try
  to provide a exhaustive bibliography. On the other hand, since this
  paper focuses on fundamental concepts of statistical dependence, a
  considerable effort has been expended to make the relevant part of
  the bibliography complete.}. In these problems, one measures
(simultaneous or serial) values of expressions of genes under
different conditions and treats them as samples from a joint
probability distribution (PD). The goal is to infer the genetic
network based on statistical dependencies in this PD.

This involves a conceptual and a technical problem. First,
surprisingly, even now there is still no agreement on what the
dependence, the interaction, is in a multivariate setting.  Instead of
a universal definition, standard statistical methods
\cite{agresti-90,joe-97} have produced a multitude of dependence
concepts applicable in restricted contexts, such as normal noise,
binary, bivariate, or metric data, etc.  Of these, the notion of {\em
  conditional} (in)dependence in the
form of {\em Bayesian networks} (BN) \cite{
friedman-04} has
proved to be very useful in biological applications. However, it is
insufficient to deal with regulatory loops, or to distinguish
independent vs.\ cooperative regulation of a gene by others (see
below).  Further, statistical dependence is a symmetric property
\cite{ku-kullback-68}, while graphs of BNs are directed.  Thus, to
infer interaction networks, we must first carefully define what we
mean by multivariate statistical dependence.

The goal is to partition the overall statistical dependence, that is,
the deviation of the joint PD (JPD) from the product of its marginals,
into contributions from interactions of different kinds (among pairs
of variables, triplets, etc.), and, better yet, from specific
combinations of variables within a kind. Many ideas have been
suggested \cite{joe-97,schlather-tawn-03,goodman-johnson-04}, but the
most natural approach is to quantify the new knowledge that comes from
looking at a complete JPD vs.\ its approximations under various
independence assumptions. For example, in contingency tables analysis,
one studies deviations of the number of observed counts from their
expectations under such assumptions
\cite{lancaster-51,roy-kastenbaum-56,darroch-62,lewis-62,mosteller-68}.
Such discussion is limited to categorical data and, importantly,
confounds the definition of dependence with sampling issues.
Information theory \cite{sw-49,ct-91} provides tools to treat
continuous and categorical data uniformly
\cite{ireland-kullback-68,kullback-68} and to generalizes bivariate
dependence measures to multivariate cases based on distributions
rather than counts.  However, some suggested information theoretic
measures
\cite{mcgill-54,garner-mcgill-56,watanabe-60,good-63,han-78,studeny-vejnarova-98,matsuda-00,bell-02,chechik-etal-02,bezzi-etal-02,schneidman-neuro-etal-03}
are not necessarily nonnegative, or involve averaging logarithms of
fractional powers of PDs. Thus they cannot characterize sizes of
typical sets and have not become universally accepted.  Instead, one
notices that the maximum entropy (MaxEnt) distributions
\cite{janes-57,lewis-59} consistent with some marginals of the JPD
introduce no statistical interactions beyond those in the said
marginals. Thus the JPD can be compared to its MaxEnt approximations
to separate dependencies included in the low order statistics from
those not present in them \cite{good-63,ku-kullback-68}.  This
approach completely characterizes multivariate interactions for binary
\cite{soofi-92,martignon-etal-00} or exponential \cite{amari-01}
distributions.  In general, it led to a definition of {\em connected}
interactions of a given order, that is, the interactions that need, at
least, the full set of marginals of this order to be captured
\cite{schneidman-etal-03}.  However, there has been no successful
attempt of defining {\em dependence among variables}, that is,
localizing (connected) interaction to particular covariates.

The second, technical, problem emerges if we agree on the measures of
dependence. The usual approach then is to use the experimental data to
infer the JPD in question and evaluate these measures for
it\footnote{We deliberately remain vague about the identity of the
  covariates. These can be the simultaneous gene expressions, in which
  case there is a big leap between inferring statistical dependencies
  and reconstructing the network.  These also can be the time lagged
  expressions, or their whole time courses, which makes the
  reconstruction easier, or something altogether different.}.
Unfortunately, the distributions may be severely undersampled.  To
fight this, the field has focused on a simplistic approach
\cite{barash-friedman-02,wiggins-nemenman-03}: (a) assume some
dependence structure; (b) regularize the JPD consistently with it and
learn it from the data; (c) evaluate the quality of fit; (d) repeat
until the dependence structure with the best fit is found.  Validity
of such analysis is sensitive to the choice of the regularization, and
a bad choice may lead to misestimation of dependencies.  As a rule,
inferring a complicated object (the JPD) in order to only find its
simple property (the dependencies) is not a good idea
\cite{vapnik-98}, and a {\em direct} estimation of dependencies
without learning the JPD is preferred\footnote{Direct estimation of
  the quantity of interest without the intermediate inference of the
  underlying PD has been useful in various contexts, in particular for
  estimating entropies \cite{nsb,nbr-04}.}.

From Shannon's work \cite{sw-49} and its later developments
\cite{ku-kullback-68,schneidman-etal-03} we know that we have to look
at information theoretic quantities to measure dependence.  Many of
these are differences of (marginal, conditional) entropies of the JPD.
While earlier works \cite{mcgill-54,ku-kullback-68} relied on good
sampling, we now know that, at least under some conditions, entropy
may be estimated reliably even when inferences about details of the
underlying PD are impossible\footnote{\label{foot:ent}The reader is
  referred to \cite{beirlant-etal-97} and to
  \url{menem.com/~ilya/pages/NIPS03} for overviews.}.
Thus the direct estimation of dependencies has a chance even for
undersampled problems.

In this paper we deal with both the conceptual and the technical
problem. First, expanding on \cite{ku-kullback-68,schneidman-etal-03}
(see also \cite{nemenman-tishby-04} for axiomatization), we
systematically characterize dependencies among variables.  Second,
we apply direct entropy estimation methods to undersampled synthetic
data to show that interactions can be uncovered even in that regime.

\section{Definitions}
Suppose we have a network of $M$ covariates $X_i$, $i=1,\dots,M$,
(called {\em nodes}, {\em expressions}, or {\em genes}) that take
random values $x_i$ respectively with the joint probability $P({\bf
  x})$.  The total statistical dependence among the variables is given
by their {\em multiinformation}, that is, the Kullback--Leibler
divergence between the JPD and the product of the marginals
\cite{schneidman-etal-03,nemenman-tishby-04}\footnote{We remind the
  readers that the Kullback--Leibler divergence, $D_{\rm KL}[P||Q] =
  \sum_x P(x) \log_2 P(x)/Q(x)$ is a natural information--theoretic
  measure of dissimilarity between PDs. It is nonsymmetric,
  nonnegative, and it is zero iff $P=Q$ \cite{ct-91}.},
\begin{multline}
  I[P] := \sum_{\{x_i\}} P({\bf x})\log_2 \frac{P({\bf x})} {\prod_i
    P(x_i)}\nipsprl{}{\\} = \left< \log_2 \frac{P({\bf x})}{\prod_i
      P(x_i)}\right>= \sum_i S[X_i] -S[{\bf X}]\,,
\end{multline}
where $\sum$'s represent summations for discrete and integration for
continuous variables respectively\footnote{In this, work we do not aim
  at mathematical rigor of the measure theoretic information theory.
  In particular, we assume that all quantities of interest exist for
  all distributions considered.}, and $S[X] := - \sum P(x) \log_2
P(x)$ is the entropy of $P(x)$.

Following \cite{schneidman-etal-03}, we localize these bits of
dependence to statistical interactions of different orders. If all
$m$-way marginals, $P(x_{i_1},x_{i_2},\dots,x_{i_m})$, are known, then
one finds an approximation to the JPD that respects the marginals, but
makes no additional assumptions about the JPD\footnote{All JPDs
  constrained by the same marginals are said to form a Fr\'echet class
  \cite{joe-97}. For metric variables and simple constraints, these
  classes are well studied. We know parametric forms
  for some of them, can check if the constraints are compatible, and
  if they determine the JPD uniquely.}. This is given by the MaxEnt,
or minimum multiinformation, problem
\cite{janes-57,ku-kullback-68,schneidman-etal-03}:
\begin{multline}
   P^{(m)} := \arg \max_{P', \{\lambda\}} \bigg\{S[P']
  \nipsprl{}{\\}  - \sum_{i_1<\cdots<i_m}^M\sum_{x_{i_1}\dots x_{i_m}}
  \lambda_{i_1\dots i_m}   (P'_{i_1\dots i_m}- P_{i_1\dots i_m})\bigg\}, \label{eq:maxentconn}
\end{multline}
where $\lambda$'s are the Lagrange multipliers enforcing the marginal
constraints; different $\lambda$'s are distinguished by their
arguments. The arguments of all functions are listed in their lower
indices, e.~g., $P_1:=P(x_1)$.  No indices means dependence on all
variables, while ${\not i}$ on all, but $X_i$.  Further, a
distribution with lower indices is a marginal, e.\ g., $P_{12} :=
\sum_{\not x_{1}\not x_{2}} P $.

A solution of any MaxEnt problem with marginal constraints has a form
of a product of terms depending on the constrained variables
\cite{csiszar-75}. In particular, for Eq.~(\ref{eq:maxentconn}),
\begin{eqnarray}
   P^{(m)} &=& \prod_{i_1<\cdots<i_m} f_{i_1\dots i_m},\;\; f\ge0,\\
   P_{i_1\dots i_m}^{(m)} &=&  P_{i_1\dots i_m},\;\; \forall
  \{i_1,\dots,i_m\},
\end{eqnarray}
where $f$'s, again distinguished by their arguments, are to be found
from the constraints.  Note the Boltzmann machine or Markov network
structure of this MaxEnt distribution
\cite{hinton-sejnowski-86,friedman-04}. In general, no analytical
solution for $f$'s exists.  However, an algorithm called the {\em
  iterative proportional fitting procedure} (IPFP)
\cite{deming-stephan-40}, which iteratively adjusts a trial solution
to satisfy each of the constraints in turn, converges to the true
solution
\cite{
csiszar-75}.

Finding $P^{(m)}$ and $P^{(m-1)}$ defines connected
information \cite{schneidman-etal-03}
\begin{eqnarray}
  I^{(m)}[P] &:=& \left< \log_2\frac{P^{(m)}}{P^{(m-1)}}
  \right>\,;\\
I[P] &=& \sum_{m=2}^M  I^{(m)}[P]\,,\label{eq:Iconsum}
\end{eqnarray}
which measures the amount of statistical interactions accounted for by
$m$-way, but not by $m-1$-way marginals. This is similar to connected
correlation functions or cumulants.

In the same spirit, to determine if a particular $m$--way interaction
contributes to $I$, we may check if fixing the corresponding
$P_{i_1\dots i_m}$ recovers any dependencies not already contained in
a {\em reference MaxEnt distribution} $Q^{*(i_1\dots i_m)}$
constrained by some other marginals. That is, we define the {\em
  interaction multiinformation}
\begin{equation}
\Delta^{(i_1\dots i_m)}:=\left< \log_2
  \frac{Q^{(i_1\dots i_m)}}{Q^{*(i_1\dots i_m)}}\right>
=   I^{(i_1\dots i_m)} - I^{*(i_1\dots i_m)},
\label{eq:delta}
\end{equation}
where $Q^{(\cdot)}$ is the {\em interaction MaxEnt distribution}
satisfying all constraints of $Q^{*(\cdot)}$ and additionally having
$Q^{(\cdot)}_{i_1\dots i_m}= P_{i_1\dots i_m}$. $I^{(\cdot)}$ and
$I^{*(\cdot)}$ are multiinformations of $Q^{(\cdot)}$ and
$Q^{*(\cdot)}$ respectively. By positivity of the Kullback--Leibler
divergence, $\Delta^{(\cdot)}\ge0$.  Thus if $\Delta^{(\cdot)}>0$,
accounting for the marginal $P_{i_1\dots i_m}$ recovers more
multiinformation, and we say that the {\em corresponding interaction
  or dependence is present with respect to $Q^{*(\cdot)}$}.
  
The problem is that $\Delta^{(\cdot)}$ depends on the choice of
$Q^{*(\cdot)}$.  To test the null hypothesis of no dependencies, we
must select the reference $Q^{*(\cdot)}$ that minimizes the
interaction information. This guarantees that interactions are
accepted {\em only} if they cannot be reduced to some other
statistical dependencies in the network.  According to
Thm.~\ref{thy:main} (see Appendix), such reference distribution is
given by
\begin{eqnarray}
  Q^{*(i_1\dots i_m)} &=& f_{{\not i}_1}\cdots f_{{\not i}_m},\;\; f\ge 0\label{eq:ref} \\
  Q^{*(i_i\dots i_m)}_{{\not i}_k} & = & P_{{\not i}_k},\; \forall
  k=1\dots m.
\end{eqnarray}
This $Q^{*(\cdot)}$ preserves all marginals of the original JPD
except those that involve all $m$ co--variates being examined for an
interaction.  This is similar to the Type III Sum of Squares ANOVA for
testing significance of predictors. In fact, since $D_{\rm KL}$ is
equal to $\chi^2$ asymptotically, the similarity is not accidental.
Dependence defined by this choice of $Q^{*(\cdot)}$ is a {\em
  generalization of the conditional dependence} with the rest of the
network as a condition.

The interaction PD, which additionally preserves the joint statistics
of $X_{i_1},\dots,X_{i_m}$, but nothing extra, is
\begin{eqnarray}
  Q^{(i_1\dots i_m)} &=& f_{{\not i}_1}\cdots f_{{\not i}_m}
  f_{i_1\dots i_m},\;\; f\ge 0 
  \label{eq:int}\\
  Q^{(i_i\dots i_m)}_{{\not i}_k} & = & P_{{\not i}_k},\; \forall
  k=1\dots m,\\
  Q^{(i_i\dots i_m)}_{i_1\dots i_m} & = & P_{i_1\dots i_m}.
\end{eqnarray}
Using such $Q$ and $Q^*$ in Eq.~(\ref{eq:delta}) defines {\em
  irreducible $m$--way interactions (dependencies)} among particular
$m$ variables.  We denote these dependencies graphically by edges
coming from the variables and meeting at an $m$--edge vertex, see
Fig.~\ref{fig:conn}.  This graphical notation generalizes BNs and
\cite{schneidman-etal-03}, were the only goal was to denote $m$--way
interactions among all combinations of
co-variates simultaneously.

\section{Examples and properties}

We consider a few examples for $M=3$ (larger $M$ is analyzed
similarly). First, a regulatory cascade, or a Markov chain: $X_1\to
X_2\to X_3$, $P(x_1,x_2,x_3) = P(x_1) P(x_2|x_1) P(x_3|x_2)$.  Looking
for $X_1X_2$ dependence, we have $I^{*(12)}= I[X_1,X_3] +I[X_2,X_3]
\le I^{(12)}= I[X_1,X_2] + I [X_2,X_3]$, where the inequality is due
to the information processing inequality, and the bound is reached
only in special cases. Thus $X_1$, $X_2$ are (generically) dependent.
Similarly, $X_2$, $X_3$ are dependent; but $\Delta^{*(13)}= 0$, and
$X_1$, $X_3$ are not (even though their marginal mutual information,
induced by other interactions, is not zero).  Checking for the triplet
interactions, we find $I^{*(123)} = I[X_1,X_2] +I[X_2,X_3] =
I^{(123)}$, thus no such dependencies are present.  If now instead
$X_2$ regulates $X_1$ and $X_3$, one sees that the dependence
structure is the same.  Both networks correspond to the graph in
Fig.~\ref{fig:conn}(a).  \nipsprl{\begin{floatingfigure}[r]{2.7in}
    \setlength{\unitlength}{.25in}\vskip -8mm }{
\begin{figure}[bt]
\setlength{\unitlength}{.25in}}
\begin{picture}(12,8)
  \thicklines
  \put(1.5,4.5){\circle{.9}}
  \put(.5,6.5){\circle{.9}}
  \put(2.5,6.5){\circle{.9}}
  \put(1.25,4.35){$X_1$}
  \put(0.25,6.35){$X_2$}
  \put(2.25,6.35){$X_3$}
  \put(1.25,4.9){\line(-1,2){.55}}
  \put(0.95,6.5){\line(1,0){1.1}}
  \put(0,4.3){$(a)$}
  \put(5.5,4.5){\circle{.9}}
  \put(4.5,6.5){\circle{.9}}
  \put(6.5,6.5){\circle{.9}}
  \put(5.25,4.35){$X_1$}
  \put(4.25,6.35){$X_2$}
  \put(6.25,6.35){$X_3$}
  \put(5.25,4.9){\line(-1,2){.55}}
  \put(4.95,6.5){\line(1,0){1.1}}
  \put(6.25,6.1){\line(-1,-2){.55}}
  \put(4,4.3){$(b)$}
  \put(9.5,4.5){\circle{.9}}
  \put(8.5,6.5){\circle{.9}}
  \put(10.5,6.5){\circle{.9}}
  \put(9.25,4.35){$X_1$}
  \put(8.25,6.35){$X_2$}
  \put(10.25,6.35){$X_3$}
  \put(9.5,4.95){\line(0,1){.78}}
  \put(8.95,6.25){\line(1,-1){.56}}
  \put(10.1,6.25){\line(-1,-1){.56}}
  \put(8,4.3){$(c)$}
  \put(1.5,.5){\circle{.9}}
  \put(0.5,2.5){\circle{.9}}
  \put(2.5,2.5){\circle{.9}}
  \put(1.25,.35){$X_1$}
  \put(0.25,2.35){$X_2$}
  \put(2.25,2.35){$X_3$}
  \put(1.75,.9){\line(1,2){.55}}
  \put(1.5,0.95){\line(0,1){.78}}
  \put(0.95,2.25){\line(1,-1){.56}}
  \put(2.1,2.25){\line(-1,-1){.56}}
  \put(0,.3){$(d)$}
  \put(5.5,.5){\circle{.9}}
  \put(4.5,2.5){\circle{.9}}
  \put(6.5,2.5){\circle{.9}}
  \put(5.25,.35){$X_1$}
  \put(4.25,2.35){$X_2$}
  \put(6.25,2.35){$X_3$}
  \put(5.25,.9){\line(-1,2){.55}}
  \put(4.95,2.5){\line(1,0){1.1}}
  \put(5.5,0.95){\line(0,1){.78}}
  \put(4.95,2.25){\line(1,-1){.56}}
  \put(6.1,2.25){\line(-1,-1){.56}}
  \put(4,.3){$(e)$}
  \put(9.5,.5){\circle{.9}}
  \put(8.5,2.5){\circle{.9}}
  \put(10.5,2.5){\circle{.9}}
  \put(9.25,.35){$X_1$}
  \put(8.25,2.35){$X_2$}
  \put(10.25,2.35){$X_3$}
  \put(9.25,.9){\line(-1,2){.55}}
  \put(8.95,2.5){\line(1,0){1.1}}
  \put(9.75,.9){\line(1,2){.55}}
  \put(9.5,0.95){\line(0,1){.78}}
  \put(8.95,2.25){\line(1,-1){.56}}
  \put(10.1,2.25){\line(-1,-1){.56}}
  \put(8,.3){$(f)$}
\end{picture}
\caption{\label{fig:conn}Examples of dependencies
  for $M=3$.}  
\nipsprl{\end{floatingfigure}
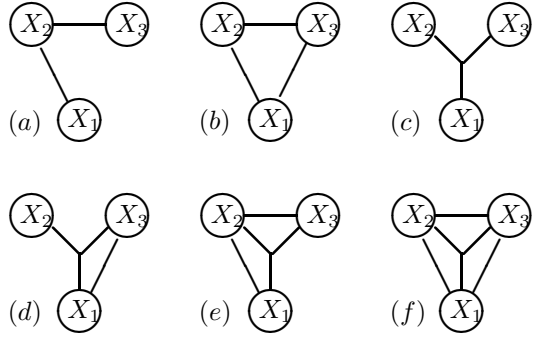} {\end{figure}}

A more interesting case is when $X_1$, $X_3$ regulate $X_2$ jointly.
Here many possibilities exist, not all of them realizable in terms of
BN modeling. First, consider independent regulation: to predict $X_2$,
one does not need to know the values of $X_1$ and $X_3$
simultaneously, $P(x_2|x_1,x_3) = f_{12}f_{23}$, e.\ g.,
$P(x_2|x_1,x_3) \propto \exp \left[-a (x_2-x_1)^2 -b(x_2-x_3)^2
\right]$ (this corresponds to {\tt OR} and {\tt AND} gates
\cite{schneidman-etal-03}, to the Lac--repressor
\cite{buchler-etal-03}, and to all regulatory models based on
independent binding of transcription factors to the DNA
\cite{reduce}).  If $P_{13}=P_1P_3$, then the dependency structure is
again as in Fig.~\ref{fig:conn}(a).  If in addition there is a
regulation $X_1\to X_3$, so that $P_{13}\neq P_1P_3$, then
$\Delta^{(13)}= D_{\rm KL}[P||Q^{*(13)}]\ge0$, and $\Delta^{(123)}=0$.
The dependency graph now has a loop in it, as in
Fig.~\ref{fig:conn}(b).

Further, in the joint regulation case one may consider a
nonfactorizable $P_{123}$ with all pairwise marginals factorizable.
An example is the {\tt XOR} gate
\cite{schneidman-etal-03,buchler-etal-03} (we were unable to construct
an explicit, normalized example for continuous variables).  In this
case, $\forall i,j,\,I^{(ij)}=0$. $\Delta^{(123)}>0$, and the
dependence structure is as in Fig.~\ref{fig:conn}(c).  Combinations of
two- and three--way dependencies are also possible
[Fig.~\ref{fig:conn}(d--f), etc.]; for example, an explicit
construction for the case (e) is $P_{123} \propto \exp
[-a(x_2-x_3x_1)^2] P_1P_3$. Such higher order dependencies are
uncommon in physics, which usually deals with low order interactions
among many variables (for example, the Hamiltonian for a spin system
is $H = -\sum_{ij} J_{ij}\sigma_i\sigma_j$; thus the JPD of spins has
no interactions of order $m>2$).  In contrast, combinatorial
regulation in genetics requires considering higher order models.

While such detailed classification is overwhelming for large $M$, some
general statements may be proven for our choice of $Q^*, Q$.  In
particular, similar to Eq.~(\ref{eq:Iconsum}), we have
\begin{equation}
  I[P] \ge \sum_{\mbox{all subsets of $\{X_i\}$}} \Delta^{(\mbox{subset})}.
\label{eq:subnet}
\end{equation}
Here the inequality is the result of our conservative approach to
identification of dependencies. To prove it, we order all $m$--way
dependencies in an arbitrary way. We then evaluate the interaction
information for the first dependency with $P^{(m-1)}$ as the reference
distribution, and take the interaction PD for each dependency as the
reference PD for the next one. Summing all $m$--way interactions gives
$I^{(m)}$, and summing over $m$ results in $I[P]$ [cf.\ 
Eq.~(\ref{eq:Iconsum})]. On the other hand, according to
Thm.~\ref{thy:main}, $\Delta^{(\cdot)}$ evaluated this way is not
smaller than the one with the references Eq.~(\ref{eq:ref}).  This
proves the above inequality.

For $M=3$, an interesting illustration of Eq.~(\ref{eq:subnet}) is
$P_{123} = P_1\,\delta(x_1-x_2) \,\delta(x_2-x_3)$.  We {\em
  correctly} identify all interactions as reducible (all $\Delta$'s
are zero). However, to account for the multiinformation in this PD,
(any) two pairwise interactions must be invoked. The degeneracy is
lifted if, for example, noise breaks the symmetry among $X$'s.

Finally, we note another interesting property of our definition. For
continuous $x_i$, the presence of interactions does not depend on
(nonsingular) reparameterizations of variables that do not mix them,
$x_i \to y_i(x_i)$ (see \cite{joe-97} for a discussion of importance
of this):
\begin{equation}
\Delta^{(\cdot)}[P(x_1,\dots,x_M)] =
\Delta^{(\cdot)}[P(y_1,\dots,y_M)].
\end{equation}
This is true since such transformations do not change factorization
properties of the MaxEnt distributions, and the Jacobians cancel in
the definition of $\Delta^{(\cdot)}$. Note that, while the definition
is reparameterization invariant, inference of distributions cannot be
done in a covariant way \cite{holy-nemenman-02}. The invariance
vanishes if instead of $P$ only a sample from it is given.


\section{Inferring networks from data}

\newlength{\figwidth}
\nipsprl{
  \setlength{\figwidth}{1.67in}
}{
  \setlength{\figwidth}{2.25in}
}
\begin{figure*}[t]
  \centerline{\parbox{\figwidth}{\includegraphics[width=\figwidth]{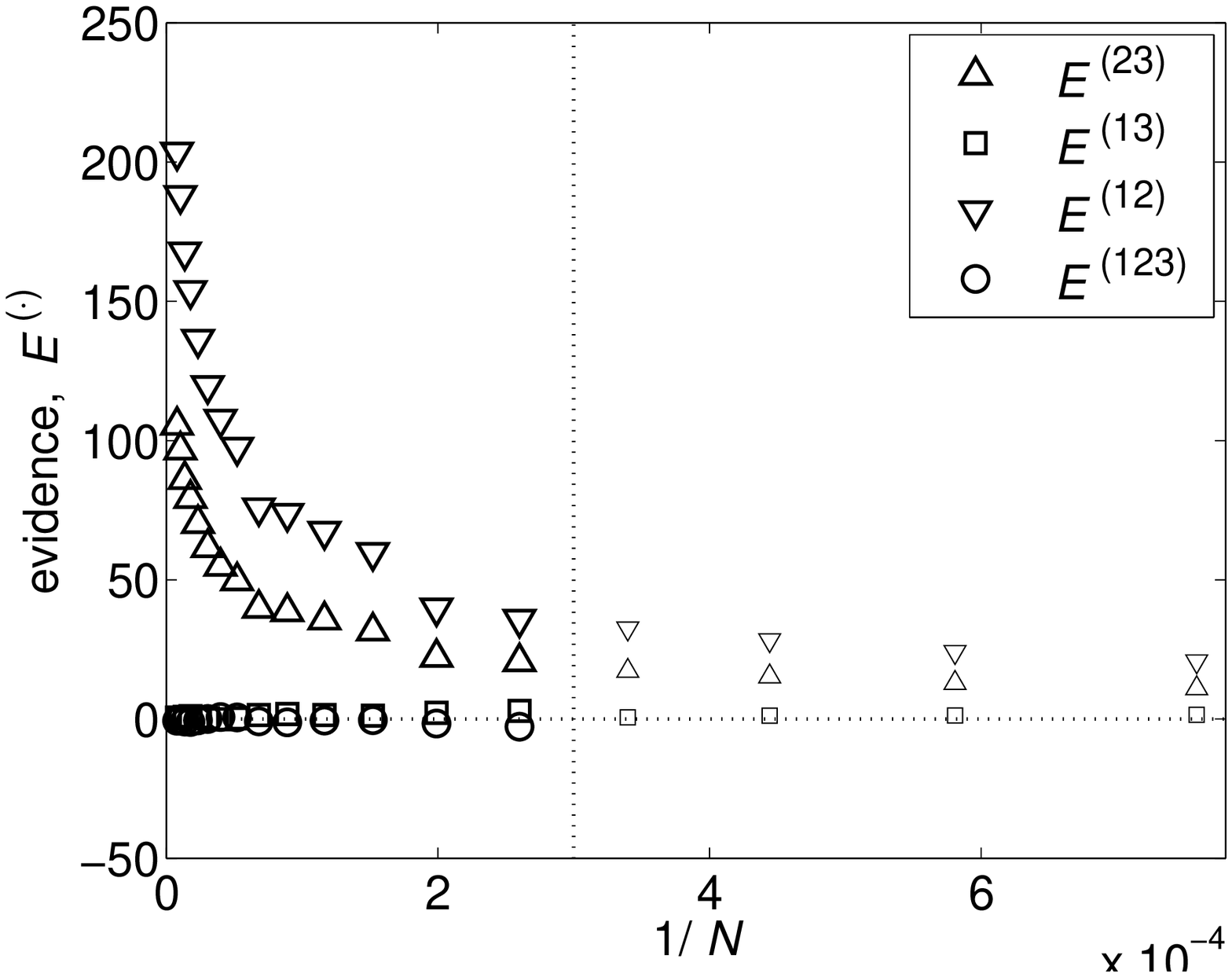}\\%
      \centerline{(a)}}
    \parbox{\figwidth}{\includegraphics[width=\figwidth]{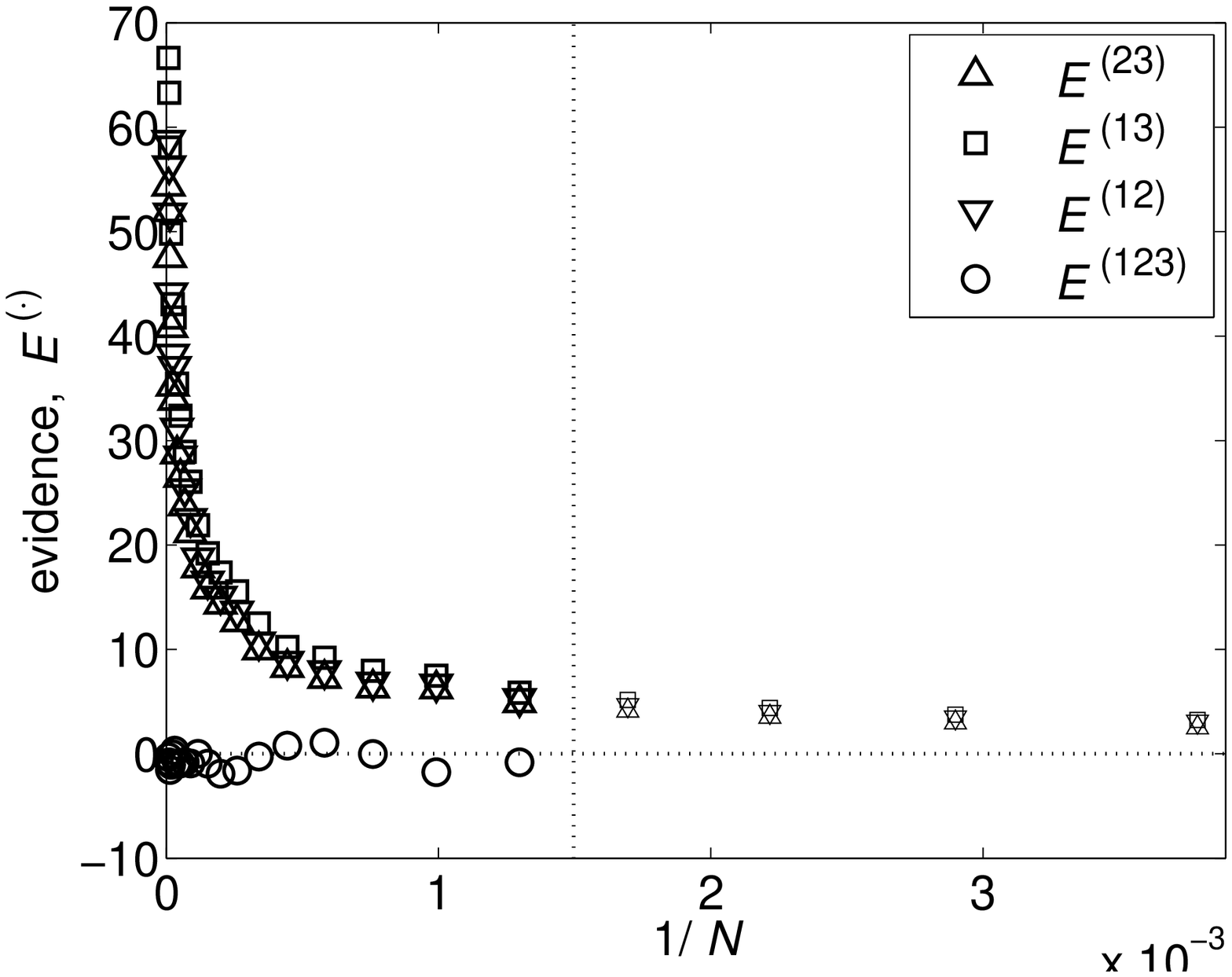}\\%
      \centerline{(b)}}
    \parbox{\figwidth}{\includegraphics[width=\figwidth]{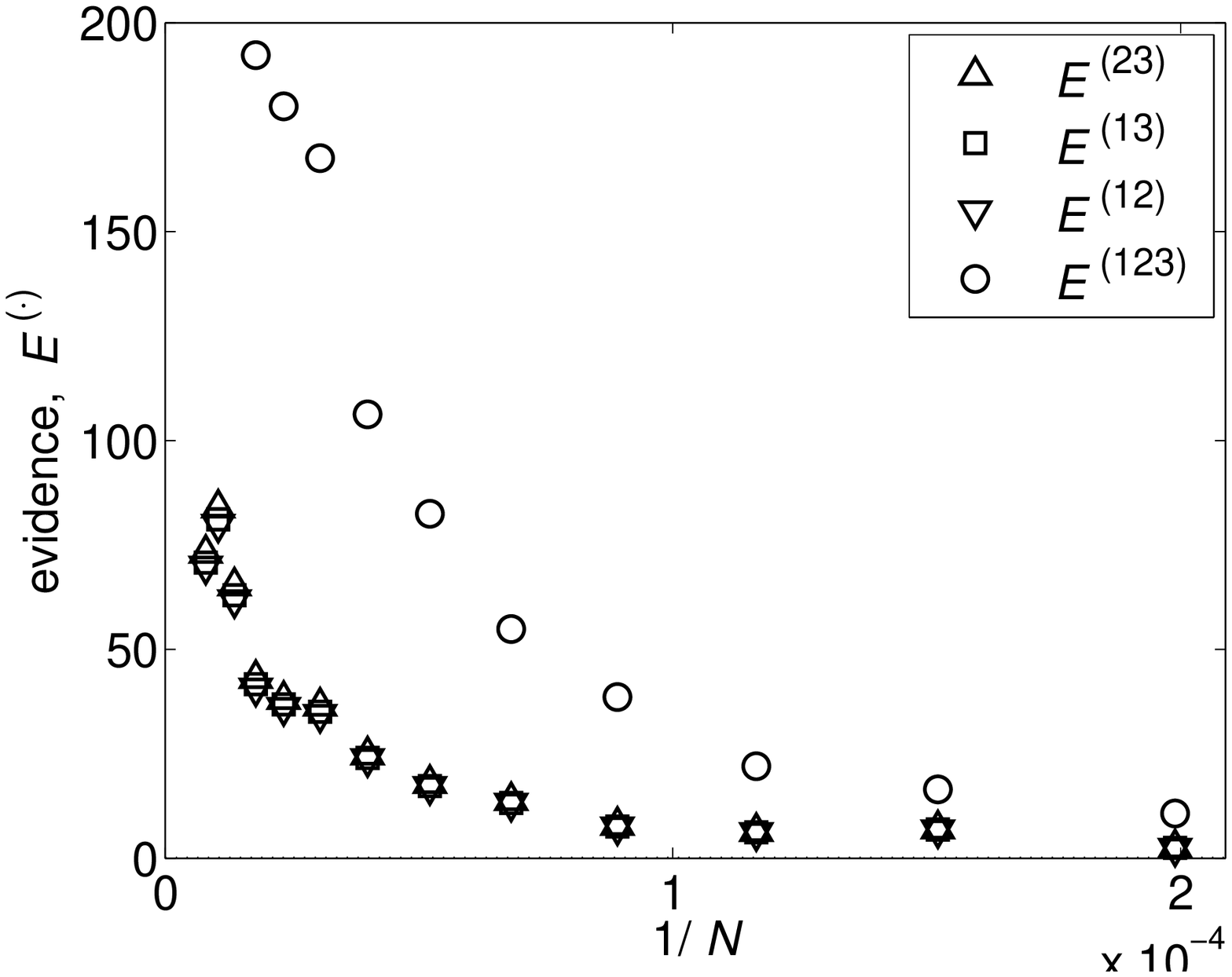}\\%
      \centerline{(f)}}}
\caption{\label{fig:infer}Inferring regulatory networks from $N$
  samples.  We used the NSB \cite{nsb} method to estimate the
  entropies (with error bars) of the JPD and its marginals directly.
  The method does not work for $Q^{*(123)}$.  Thus IPFP was applied to
  the counts and the entropy $S^{*(123)}$ of the solution was
  evaluated and extrapolated for $1/N\to0$ following Strong et al.\ 
  \cite{strong-98} to account for the sample size dependent bias.  The
  statistical error for each sample size $N$ was determined by
  bootstrapping, and the resulting extrapolation error was used for
  $\delta S^{*(123)}$.  This approach works since the MaxEnt
  constraints, like those in Eq.~(\ref{eq:maxentconn}), are linear in
  the unknown JPD $P$, making the biases of $S[P]$ and $S^{*(123)}$
  behave similarly.  Finally, $\Delta^{(\cdot)}$ were calculated as
  the differences of the appropriate entropies, and
  $\delta^2\Delta^{(\cdot)}$ as the sums of squares of the entropy
  errors.  (a) Network in Fig.\ref{fig:conn}(a).  To the left of the
  vertical dotted line, $N\approx 3000\gtrsim2^{S^{*(123)}}\ll
  K\approx125000$, the sample size corrections are reliable, and all
  entropies are known well.  There is evidence only for $X_1X_2$ and
  $X_2X_3$ interactions, just as it should be.  For smaller $N$, the
  method of Strong et al.\ fails, but NSB works until $N\sim
  2^{1/2S[P]}\approx 60$.  For pairwise interactions, we may replace
  $S^{*(123)}$ by $S[P]$ (denoted by smaller markers on the plot) and,
  since $E^{(13)}$ stays zero nonetheless, and
  $I[X_1,X_2]+I[X_2,X_3]=I[P]$, we still recover the interactions
  correctly.  (b) Network in Fig.~\ref{fig:conn}(b).  Again, to the
  left of the line, $N\approx700\gtrsim2^{S^{*(123)}}$, all entropies
  are determined reliably, and there is evidence for all three
  pairwise interaction, but not for the triplet interaction. To the
  right of the line, NSB still works, but now we cannot disentangle
  the loop from the three--way dependence without estimating
  $S^{*(123)}$.  (c) Network in Fig.~\ref{fig:conn}(f).  Only the
  regime $N\gtrsim2^{S^{*(123)}}\approx5000$ is shown. The evidence
  for all three pairwise interactions and for the triplet interaction
  is barely significant for small $N$ but grows fast.}
\end{figure*}

A big advantage of our definition of statistical dependencies in terms
of the MaxEnt approximations is that it can be applied even when the
underlying PDs are undersampled, and the corresponding factorizations
cannot be readily observed. For $K$, the cardinality of a
variable\footnote{In genomics, continuous expression levels are
  routinely discretized into three states: up, down, and baseline.
  Thus we decided to focus on the discrete case in view of its
  relevance and conceptual simplicity.  Measuring dependencies for
  continuous variables follows a similar route, with the estimation of
  entropies performed by one of the many methods reviewed in
  \cite{beirlant-etal-97}.}, larger than the number of samples, $N$,
we cannot estimate the PDs reliably, but entropic quantities, and,
therefore, the interactions are inferable\footnotemark[4]
(some progress is possible even for $N\sim \sqrt{K}$)%
. To show this, we used Dirichlet priors \cite{nsb} to
generate random probability distributions with different interaction
structures, $M=3$, and with marginal cardinalities $K_i\approx50$.  We
generated random samples of different sizes, $N=50\dots125000$ from
these distributions and tested the quality of inference of the
dependencies as a function of $N$.  To measure it, we used the {\em
  evidence} for some dependency, $E^{(\cdot)} :=
\Delta^{(\cdot)}/\delta \Delta^{(\cdot)}$, where $\delta
\Delta^{(\cdot)}$ is the statistical error of the interaction
multiinformation.  If $E^{(\cdot)}$ is large, the dependency is
present.  According to Fig.~\ref{fig:infer}, proper recovery is
possible for $N\ll K=K_1K_2K_3$ {\em with few assumptions} about the
form of the PDs.

With modern entropy estimation techniques \cite{nsb}, our approach
will work even for severely undersampled JPD. The bottleneck is the
estimation of the maximum entropy consistent with the marginals, which
currently requires substantial sampling of the marginals
\cite{strong-98}. This is encouraging, since they may be well sampled
when the JPD is not. However, it is still essential to develop
techniques to infer maximum entropies directly.  Further, the
interaction information is the difference of entropies.  It may be
small when its error, which is a quadratic sum of the entropy errors,
is large.  This leads to uncertainties about dependencies even for
reliably estimated entropies, see the small $N$ region in
Fig.~\ref{fig:infer}(c).  Therefore, a method that directly estimates
$\Delta$ will be preferred over another entropy--based technique.
Finally, as in Fig.~\ref{fig:conn}(a), variables may have nonzero
mutual (or higher order) information and no direct interactions. Thus,
if $X_2$ was unobserved, we would have inferred a dependence between
$X_1$ and $X_3$.  Similarly, spurious higher order interactions may
also emerge.  Our method, just like most other assumption--free
methods, may fail for hidden variables.

For genomic applications, the number of different expression
measurements is $N\lesssim100$, and it is not nearly enough to
estimate $\Delta$'s and to infer the full interaction network of, say,
$M\approx6000$ genes in a yeast.  However, for ternary discretization
of expressions, with the Strong et al.\ entropy estimation, one will
not find significant evidence for $I^{(m)}$ beyond $m^*\sim \log_3 N
\approx 4$ (or somewhat larger if PDs are far from being uniform).
Then one can replace $P$ by $P^{(m^*)}$ in Eq.~(\ref{eq:ref}) and
study interactions up to the order $m^*$ with respect to this
JPD.  It is possible that most interactions in genomes are of
such low orders.  Additionally, if methods like NSB \cite{nsb} are
developed for MaxEnt analysis, one should be able to push for $m^*\sim
2\log_3 N \approx 8$, and this is the primary goal of our future work.

In summary, we have formalized the concept of multivariate dependence,
suggested a way to infer dependencies from data, tested the suggestion
on undersampled synthetic examples, and hinted at possible
applications to genomic research.

\nipsprl{\subsection*{Appendix}}{
  \appendix*
  \section{}}
\begin{theorem}
  \label{thy:main}
  Let $\{C\}$ be a set of noncontradictory marginal constraints and
  $Q^{{C}}$ be the MaxEnt distributions satisfying these constraints.
  Further, let $C_0$ and $C_1$ be additional constraints (possibly
  included in $\{C\}$), and $Q^{C0}$, $Q^{C1}$, and $Q^{C01}$ be the
  MaxEnt PDs satisfying $\{C\} \cup C_0$, $\{C\}\cup C_1$, and
  $\{C\}\cup C_0 \cup C_1$ respectively. Then
  \begin{equation}
    \left<\log_2\frac{Q^{C01}}{Q^{C0}}\right> \le
    \left<\log_2\frac{Q^{C1}}{Q^{C}}\right>,
  \end{equation}
  where the averaging is performed over $Q^{C01}$.
\end{theorem}
Intuitively, this says that the interaction multiinformations depend
on the order in which the interactions are considered.  Dependency
bits will be accounted for by the first marginal able to explain them,
attributing less bits to later constraints. At present, this theorem
has been extensively tested by numerical simulations, but still
remains a conjecture.

\section*{Acknowledgments}
I thank W.~Bialek and A.~Califano for asking the right questions and
N.~Tishby and C.~Wiggins for stimulating discussions.  This work was
supported by NSF grants PHY99--07949 to Kavli Institute for
Theoretical Physics and ECS--0332479 to C.~Wiggins and myself.

{\nipsprl{\small}{}
\bibliographystyle{\nipsprl{aps}{unsrtnat}} \bibliography{infonet}
}

\end{document}

%% file: aliases.tex
\newcommand{\bde}{\begin{description}}

\newcommand{\ben}{\begin{enumerate}}
\newcommand{\beq}{\begin{eqnarray}}

\newcommand{\beqn}{\begin{eqnarray*}}

\newcommand{\bqu}{\begin{quote}}

\newcommand{\bvb}{\begin{verbatim}}

\newcommand{\ea}{\end{eqnarray}}

\newcommand{\ede}{\end{description}}
\newcommand{\een}{\end{enumerate}}
\newcommand{\eeq}{\end{eqnarray}}
\newcommand{\eeqn}{\end{eqnarray*}}

\newcommand{\equ}{\end{quote}}

\newcommand{\evb}{\end{verbatim}}

\newcommand{\pmbeg}{\begin{pmatrix}}
\newcommand{\pmend}{\end{pmatrix}}

